\documentclass[%
 reprint,
 amsmath,amssymb,
 aps,
 pra,
]{revtex4-1}

\usepackage{graphicx}
\usepackage{dcolumn}
\usepackage{bm}
\usepackage{float}
\usepackage{multirow}

\begin{document}

\preprint{APS/123-QED}

\title{High-speed noise-free optical quantum memory}

\author{K.~T.~Kaczmarek$^{1*}$, P.~M.~Ledingham$^1$, B.~Brecht$^1$, S.~E.~Thomas$^{1,2}$, G.~S.~Thekkadath$^{1,3}$, O.~Lazo-Arjona$^1$, J.~H.~D.~Munns$^{1,2}$, E.~Poem$^{4}$, A.~Feizpour$^1$, D.~J.~Saunders$^1$, J.~Nunn$^{5}$, I.~A.~Walmsley$^{1}$}
\email{krzysztof.kaczmarek@physics.ox.ac.uk\\ i.walmsley1@physics.ox.ac.uk}

\affiliation{%
$^1$Clarendon Laboratory, University of Oxford, Parks Road, Oxford, OX1 3PU, UK.\\
$^2$QOLS, Blackett Laboratory, Imperial College London, London SW7 2BW, UK. \\
$^3$University of Ottawa, 25 Templeton St, Ottawa, K1N 6N5, Canada.\\
$^4$Department of Physics of Complex Systems, Weizmann Institute of Science, Rehovot 7610001, Israel.\\
$^5$Centre for Photonics and Photonic Materials, Department of Physics, University of Bath, Claverton Down, Bath BA2 7AY, UK.}%

\date{\today}

\begin{abstract}
Optical quantum memories are devices that store and recall quantum light and are vital to the realisation of future photonic quantum networks. To date, much effort has been put into improving storage times and efficiencies of such devices to enable long-distance communications. However, less attention has been devoted to building quantum memories which add zero noise to the output. Even small additional noise can render the memory classical by destroying the fragile quantum signatures of the stored light. Therefore noise performance is a critical parameter for all quantum memories. Here we introduce an intrinsically noise-free quantum memory protocol based on two-photon off-resonant cascaded absorption (ORCA). We demonstrate successful storage of GHz-bandwidth heralded single photons in a warm atomic vapour with no added noise; confirmed by the unaltered photon number statistics upon recall. Our ORCA memory meets the stringent noise-requirements for quantum memories whilst combining high-speed and room-temperature operation with technical simplicity, and therefore is immediately applicable to low-latency quantum networks.
\end{abstract}

\maketitle

\section{Introduction}
Light is the ideal information carrier for a future quantum internet \cite{Kimble2008}, as its properties are not degraded by noise in ambient conditions, and it can support large bandwidths enabling fast operations and a large information capacity. The quantum internet will most likely be comprised of a large-scale distribution of nodes --- small networks made of composite systems that process photonic quantum information --- coupled together via long haul interconnects. Such quantum networks promise to revolutionise computing, simulation, and communication. Quantum memories, devices that store, manipulate, and release on demand quantum light, have been identified as crucial components for each network element, because they facilitate scalability. This has motivated diverse research efforts on many fronts, with fast \cite{England2015, Fisher2017}, long-lived \cite{Heinze2013, Jobez2015, Seri2017}, efficient \cite{Hedges2010, Chen2013, Cho2016}, single- and multi-mode \cite{Nunn2007, Usmani2010, Gundogan2015, Laplane2016} optical memories and light-matter processors \cite{Hosseini2009, Reim2012, Maxwell2013, Fisher2016, England2016} being demonstrated. However, regardless of the unique applicability of each memory technology within a quantum internet, there is an additional overarching requirement --- the memory must be noise-free.

A quantum memory may be considered to be noise free if both the mean number of photons added by the memory and the variance of the added photon number are small. Ideally these quantities would remain unchanged from the input to the output of the memory. This can be verified by measuring the normalised Glauber correlation functions \cite{Glauber1963}, in particular the heralded auto-correlation $g^{(2)}_\mathrm{h}$ of the input and recalled light. It is important to note that it is insufficient to predict guaranteed quantum operation by only measuring the mean of the noise, because even a very small average amount of noise \cite{Namazi2017} can impair quantum signatures if the variance of the noise is large, e.g. thermal \cite{Michelberger2015}. To date, preservation of photon number statistics upon recall has only been demonstrated in narrowband atomic quantum memories \cite{Chaneliere2005,Eisaman2005,Zhou2012,Ding2016}. These are not compatible with high-speed photonic networks, such as classical optical communication networks that operate at gigahertz rates. Quantum photonic networks could inherit such high operational rates, but to date no quantum memory has demonstrated high-speed compatibility with the required zero-noise operation.

\begin{figure*}[t]
\includegraphics[width=\textwidth]{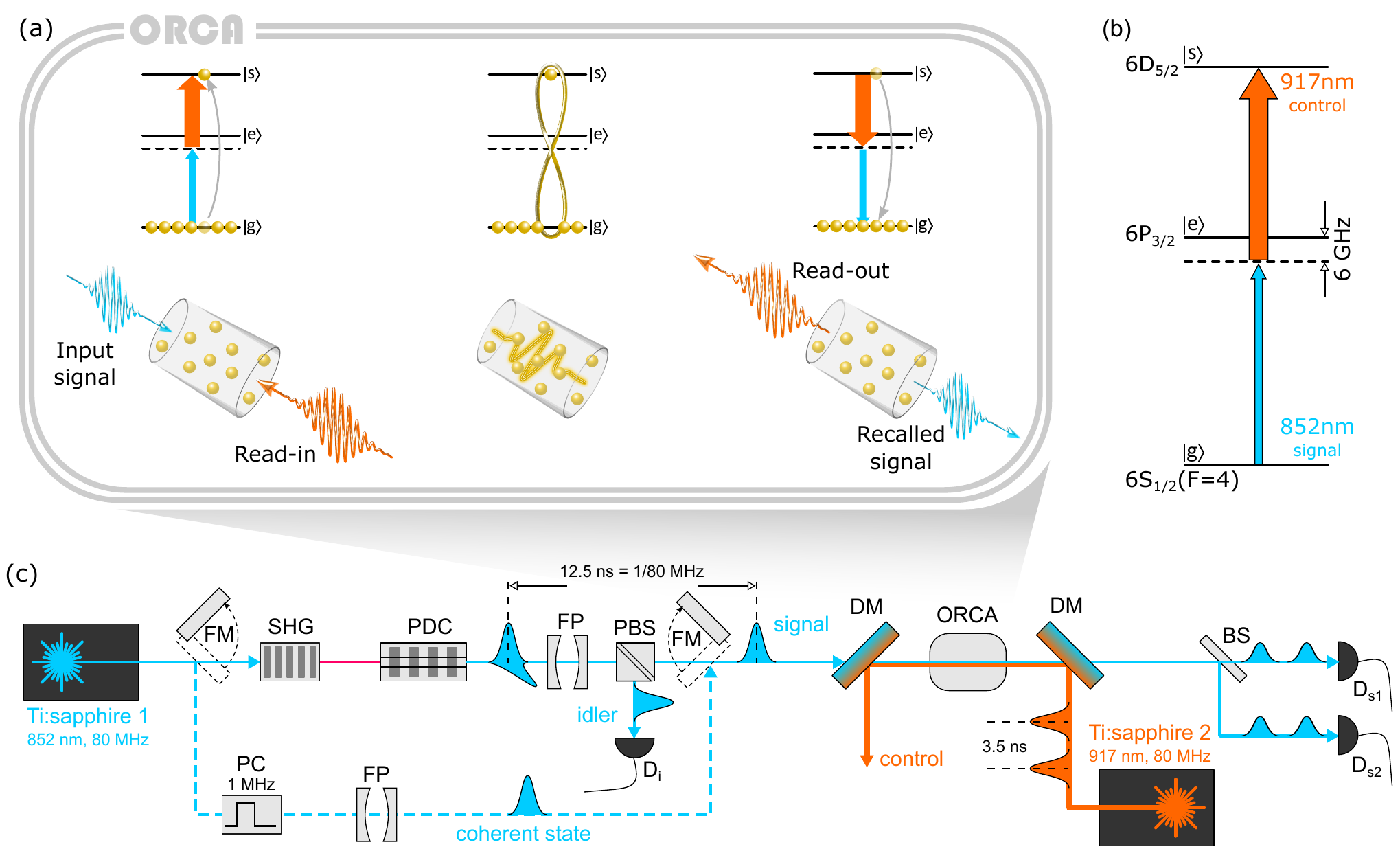}
\caption{\textbf{The ORCA protocol and experimental setup.}
(a) The ORCA protocol. (left) Storage: a weak input signal pulse and strong read-in control pulse are overlapped counter-propagating in an atomic vapour. The broadband fields are on two-photon resonance with a doubly-excited state $|s\rangle$, while being far-detuned from the intermediate state $|e\rangle$. (center) The storage maps the input signal to a collective atomic coherence (yellow twisted line) between the ground state $|g\rangle$ and $|s\rangle$. (right) Recall: applying a read-out control pulse after the desired storage time leads to a re-mapping of the atomic coherence back into an optical field and thus re-emission of the signal in the forward direction. 
(b) The relevant atomic levels in the current experimental implementation in warm caesium vapour. Under broadband excitation the atomic configuration can be treated to first order as a three-level system \cite{Huber2011}.
(c) Schematic of the setup (see text for more details). Ti:sapphire - mode-locked titanium sapphire laser; FM - flip mirror; SHG - second harmonic crystal; PDC - waveguide photon source; FP - Fabry-Per\'ot etalons with total transmission bandwidth of $\sim1$ GHz; PC - pulse picker; PBS - polarising beamsplitter; D$_\mathrm{i}$ - single-photon avalanche photodiode (APD) detector for the idler; DM - dichroic mirror; BS - beamsplitter; ORCA - caesium cell; D$_\mathrm{s1}$, D$_\mathrm{s2}$ - single-photon APDs for the signal.
}
\label{fig:protocol_experiment}
\end{figure*}

Here we introduce and demonstrate the Off-Resonant Cascaded Absorption (ORCA) memory protocol, which provides a viable real-world platform that does not measurably degrade the quantum character of the recalled light compared to the input, verified by measuring the photon number statistics.

\section{Off-Resonant Cascaded Absorption (ORCA) memory}

The operational principle of the ORCA memory protocol is summarised in Fig. \ref{fig:protocol_experiment} (a). ORCA utilises a three-level atomic cascade configuration, where a strong off-resonant ``control'' field mediates the mapping of an optical ``signal'' field into an atomic coherence between the ``ground'' ($|g\rangle$) and ``storage'' ($|s\rangle$) states. The fields are arranged in a counter-propagating configuration, in order to reduce motion-induced dephasing of the distributed $|g\rangle$-$|s\rangle$ quantum coherence in the warm atomic ensemble \cite{Zhao2009}. Similarly to the broadband Raman memory protocol \cite{Reim2010a}, the acceptance bandwidth of ORCA is determined by the control pulse bandwidth, although in ORCA it is not in principle limited by the ground state splitting of the atomic storage medium.

The storage state here is a doubly-excited electronic state, which has no thermal excitations even at high temperatures. Therefore the protocol in principle requires no preparation of the atomic ensemble prior to storage, and there is no contamination of the recalled fields due to imperfect optical pumping. This points towards the main feature of the ORCA memory in that it is fundamentally noise-free. The signal and control wavelengths can be chosen such that the control photons are significantly detuned from the populated transition (THz detunings are readily available in the rich atomic structure of alkalis). This effectively eliminates any control field induced scattering or fluorescence noise \cite{Raymer1977}. More importantly though, due to the cascade configuration, there is no scattering process that could populate the storage state, and so four-wave mixing noise \cite{Nunn2017}, which has so-far limited the usefulness of broadband quantum memories \cite{Michelberger2015}, is eliminated. Finally, efficient suppression of control field leakage on the output detection is readily achievable using off-the-shelf low-loss interference filters, in principle enabling \emph{external} device efficiencies approaching the \emph{internal} memory efficiency. 

As a proof-of-principle demonstration, we implement ORCA with near-infrared light in warm caesium vapour. We use the Cs D2 line at 852 nm for our signal field, with $6S_{1/2}(F=4)$ as the ground state $|g\rangle$ and the $6P_{3/2}(F=3,4,5)$ manifold as the ORCA intermediate state $|e\rangle$. A strong 917 nm control field ($\sim0.9$ GHz pulse bandwidth) couples this signal to the storage state $|s\rangle$, i.e. the $6D_{5/2}(F=2,3,4,5,6)$ manifold. We detune both fields by 6 GHz from the intermediate state towards the ground state, enabling good coupling with negligible ($<2\%$) linear absorption. We first benchmark the standard memory performance parameters with weak (mean photon number of $\langle\hat{n}\rangle_\mathrm{in}\approx{2}$) coherent signal pulses ($\sim540$ ps duration), as shown in Fig. \ref{fig:protocol_experiment} (c).

\section{Single-photon level characterisation of ORCA memory in warm Cs vapour}

\begin{figure}[h]
\centering
\includegraphics[width=\columnwidth]{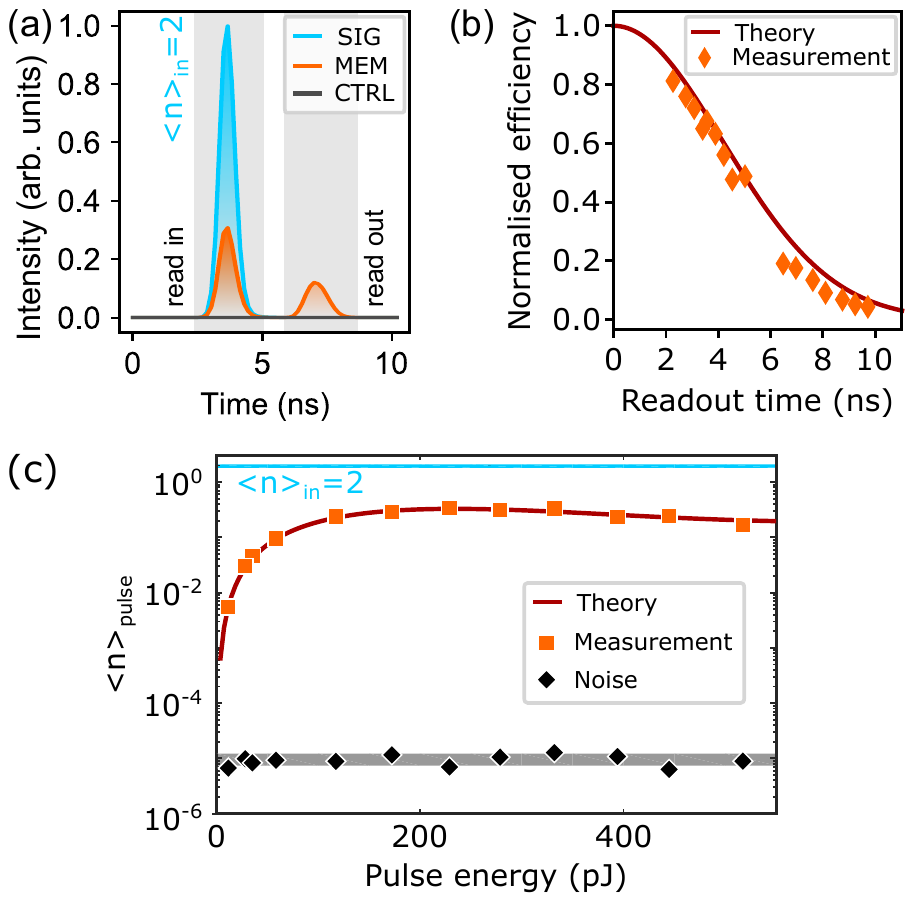}
\caption{\textbf{Classical characterisation of ORCA.}
(a) Histogram of the arrival time statistics of a weak coherent state with respect to a trigger derived from the laser system. ``SIG'' is signal on with the control field off. ``MEM'' is both signal and control fields on. ``CTRL'' has the control field on, but the signal field off. We use this measurement to determine the mean added memory noise. The memory efficiencies are obtained from the ratio of the areas under the ``SIG'' pulse and the ``MEM'' recall pulse. The temporal length of the detected signal is limited by detector jitter.
(b) Measurement of the memory lifetime (diamonds) and the prediction of our theoretical model (line).
(c) Recalled average photon number (squares) and noise (diamonds) as a function of control pulse energy for a storage time of 3.5 ns and input mean photon number of 2. Also shown is the fit of our theoretical model to the data (dark line). All error bars are smaller than the symbol size.
}
\label{fig:characterisation}
\end{figure}

Fig. \ref{fig:characterisation} (a) shows storage and recall of a single-photon level pulse. The signal absorption is approximately 70\% when the read-in control field is turned on, and about 20\% of this can be read out after 3.5 ns of storage time. The memory decoheres after some time, reducing the recall efficiency, as shown in Fig. \ref{fig:characterisation} (b). We measure a memory $1/e$-lifetime of $5.4(1)$ ns, limited by residual motion-induced dephasing in the warm ensemble and quantum interference between the different hyperfine states in the $6D_{5/2}$ manifold, i.e. our storage state.

We model the memory using a standard Maxwell-Bloch approach. The dynamics of the atomic density matrix $\hat{\rho}(v)$ in different velocity classes is solved under coupling with signal and control fields, including spontaneous emission. In order to capture the effect of hyperfine structure on memory lifetime we include the 12 atomic states corresponding to the $6S_{1/2}, 6P_{3/2}, 6D_{5/2}$ hyperfine manifolds in $\hat{\rho}(v)$. Each velocity class evolves under a Doppler-shifted Hamiltonian. The signal field is coupled to the total density matrix $\hat{\rho}=\int \mathrm{d}v g(v) \hat{\rho}(v)$ (where $g(v)$ is a Maxwell-Boltzmann velocity distribution) through the source term of Maxwell's wave equation. The control field is assumed to propagate without dispersion from the atomic vapour, since it is so far detuned from any atomic resonance involving the populated state. We numerically solve the Maxwell-Bloch equations using the experimental parameters and tabulated atomic data, with only electric dipole matrix elements and signal/control temporal overlap as free parameters. We find excellent agreement between the measurement (diamonds, Fig. \ref{fig:characterisation} (b)) and our theoretical prediction (line). This confirms that the memory coherence time in Cs is limited by Doppler broadening (due to the incomplete cancellation of the signal and control wavevectors) leading to motion-induced dephasing, emphasised by quantum interference between different hyperfine state components in the generated atomic coherence. The memory lifetime can be improved by moving to a different atomic medium (e.g. $\sim100$ ns in warm rubidium vapour \cite{SuppMat,Finkelstein2017}).

Next we measure the memory efficiency at a storage time of $3.5$ ns as a function of the control pulse energy (read-in/-out pulse energies being equal to each other), as shown in Fig. \ref{fig:characterisation} (c). The measured recalled photon number $\langle\hat{n}_\mathrm{mem}\rangle$ (squares) closely follows the theoretically expected curve (dark line). We measure a maximum memory efficiency of $\eta_\mathrm{max}=16.77(2)\%$. Including filtering and other losses between the front of the memory and the detectors, this leads to a device end-to-end efficiency of $\sim 5$\%. In the present demonstration, the efficiency was limited by the coherence time of the memory and available control pulse energy. By switching to a different atomic system (e.g. rubidium) and adjusting operation parameters such as atomic density and control pulse energies, our theoretical model predicts memory efficiencies in excess of 50\% (since gain processes \cite{Thomas2016} are absent in ORCA, we expect the noise-free properties to survive at high efficiencies).

We also measure the control-field induced noise counts $\langle\hat{n}\rangle_\mathrm{noise}$ (diamonds), which do not show any dependence on control pulse energy. We benchmark the noise performance of the memory by evaluating $\mu_1=\langle n^\mathrm{noise}\rangle/\eta$, i.e. the ratio of the average number of noise photons per control pulse $\langle n^\mathrm{noise}\rangle$ and $\eta$ \cite{Gundogan2015}. For a memory efficiency of $16.77(2)\%$, we find $\mu_1=3.8(9)\times10^{-5}$, the lowest ever reported from an atom-based quantum memory. Moreover, the detected ``noise'' is consistent with detector dark counts (grey shaded area), strongly suggesting that the memory itself generates no noise. However, we emphasise again that only a measurement of the recalled photon number statistics can confirm true quantum operation.

\section{Quantum storage of single photons in warm vapour using ORCA}

\begin{figure}[h]
\centering
\includegraphics[width=\columnwidth]{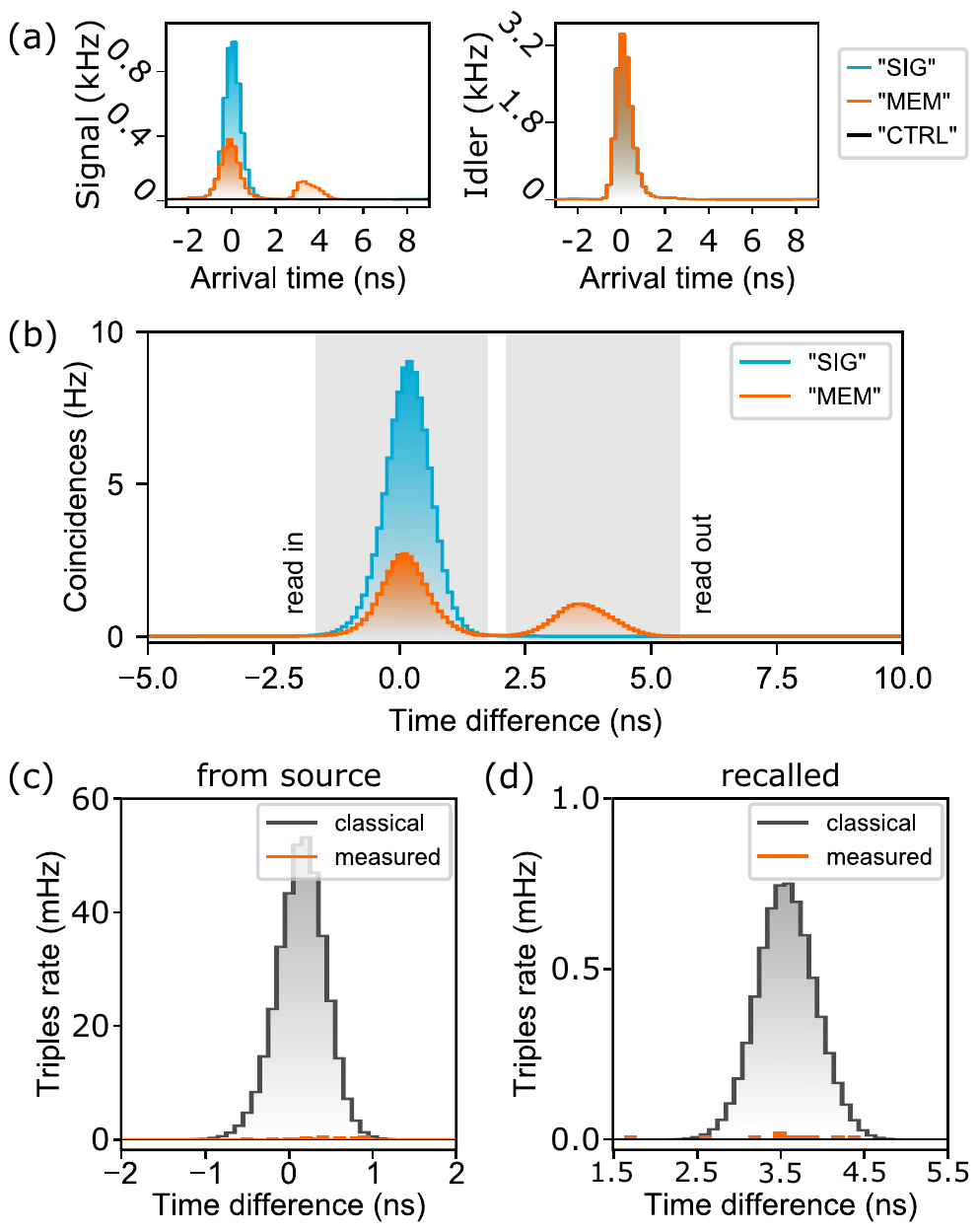} 
\caption{\textbf{Noise free single-photon storage.}
(a) Arrival time traces of accumulated D$_\mathrm{s1,s2}$ (left) and D$_\mathrm{i}$ (right) clicks with respect to an external 1 MHz trigger. Labeling is the same as in Fig. \ref{fig:characterisation}a.
(b) Histogram of the time difference between D$_\mathrm{i}$ and D$_\mathrm{s1,s2}$ coincidence clicks, with the control off (``SIG'') and on (``MEM''). The square shaded areas correspond to the 3.5 ns integration windows for storage and recall.
(c) Histogram of the time difference between D$_\mathrm{i}$ clicks, and D$_\mathrm{s1}$-D$_\mathrm{s2}$ coincidences, i.e. triple coincidence histogram (labeled ``measured'') for the ``SIG'' configuration. Also shown is the product of the two-fold D$_\mathrm{i}$-D$_\mathrm{s1}$ and D$_\mathrm{i}$-D$_\mathrm{s2}$ coincidences, normalized by the D$_\mathrm{i}$ counts, i.e. predicted triple coincidence histogram for independent coherent states of the same average photon rate as the PDC (labeled ``classical''). The ratio between the two histograms corresponds to the measured heralded auto-correlation function $g^{(2)}_\mathrm{h}$.
(d) Same as (c), but for the recalled signal in the ``MEM'' configuration. In all traces temporal resolution is limited by detector response.
}
\label{fig:quantum}
\end{figure}

To demonstrate quantum-limited operation of ORCA, we test the storage and recall of heralded single photons. These are generated by means of type-II parametric down-conversion (PDC) in a periodically poled potassium titanyl phosphate waveguide. The source produces THz-bandwidth orthogonally polarised pairs of ``signal'' and ``idler'' photons, both of which are consequently filtered down to $\sim1$ GHz bandwidth centred at the signal frequency using a series of Fabry-Per\'ot etalons and grating filters \cite{Michelberger2015}. Detecting the idler heralds the presence of a single signal photon at the memory. Our heralding efficiency before the memory is $\eta_\textrm{herald}\approx5\,\%$.

For the single photon experiment, the read-in and read-out control pulse energies were chosen to be 0.21(1) and 0.97(1) nJ respectively, and the photons are stored in the ORCA memory for 3.5 ns. Owing to the short lifetime, and without the need for time-consuming atomic state preparation, we are able to operate the single-photon experiment at the full 80~MHz repetition rate of our PDC pump, greatly increasing the rate at which we can acquire photon statistics. We use a Hanbury-Brown-Twiss detection setup, as shown in Fig. \ref{fig:protocol_experiment} (c), to reconstruct the quantum photon number statistics of the stored/recalled signal fields. Fig. \ref{fig:quantum} (a) shows the photon arrival time traces for both signal and idler. As was the case during the ``classical'' characterisation, we do not see any control field induced noise counts in the ``CTRL'' traces. For PDC photons, we measure the memory efficiency (storage and recall) to be $\eta=14.6(1.9)\%$, close to that of the weak coherent state signal (the difference being attributed to a slight bandwidth mismatch).

We investigate the quantum operation of our memory by measuring Glauber correlation functions. Fig. \ref{fig:quantum} (b) shows the detected coincidence clicks between the detectors $\mathrm{D_{i}}$ \& $\mathrm{D_{s1/2}}$ at different times with the control off (``SIG'') and on (``MEM''). First, we evaluate the cross-correlation function $g^{(1,1)}$ of signal and idler photons. $g^{(1,1)}$ is defined as $p_\mathrm{si}/p_\mathrm{s}p_\mathrm{i}$, where $p_\mathrm{si}$ is the probability for a signal-idler coincidence click, and $p_\mathrm{s(i)}$ is the signal (idler) click probability. Values of $g^{(1,1)}>2$ signify non-classical correlations \cite{Farrera2016}. We measure $g^{(1,1)}=130(5)$ for the input signal field, and upon recall obtain $g^{(1,1)}=120(5)$, clearly exceeding the classical bound and demonstrating the preservation of non-classical correlations in ORCA. We attribute the slight reduction of the mean value $g^{(1,1)}$ in the read-out due to increased dark count contamination.

Finally, we demonstrate that ORCA preserves the photon number statistics of our input signal. To this end, we evaluate the heralded auto-correlation function $g^{(2)}_\mathrm{h}$; related to the photon number variance \cite{mandel1979}. The heralded auto-correlation is defined as $g^{(2)}_\mathrm{h}=p_\mathrm{(s1,s2|i)}/p_\mathrm{(s1|i)}p_\mathrm{(s2|i)}$. Here, $p_\mathrm{(s1,s2|i)}$ is the conditional probability of detecting a coincidence between $\mathrm{D_{s1}}$ and $\mathrm{D_{s2}}$ given a click in $\mathrm{D_{i}}$, and $p_\mathrm{(s1|i)}$ ($p_\mathrm{(s2|i)}$) is the probability to detect a click in $\mathrm{D_{s1}}$ ($\mathrm{D_{s2}}$) given a click in $\mathrm{D_{i}}$. A $g^{(2)}<1$ indicates sub-poissonian photon statistics, with lower variance than classical light. We measure the $g_\mathrm{h}^{(2)}$ of our input signal (Fig. \ref{fig:quantum} (c)) to be $0.020(5)$, confirming that we herald high-quality single photons which are a very sensitive probe for assessing noise performance \cite{Michelberger2015}. Upon recall (Fig. \ref{fig:quantum} (d)), we obtain $g_\mathrm{h}^{(2)}=0.028(9)$. Within our measurement accuracy we observe no change in $g_\mathrm{h}^{(2)}$, which proves that the memory adds zero noise.

\section{Conclusion}

In conclusion, we have introduced and demonstrated a noise-free atomic quantum memory --- the ORCA memory --- which operates at ambient conditions and is compatible with broadband light. We have furthermore characterised the memory performance, and developed a complete theoretical model of the experiment which describes our data well. Using this model, we expect the performance of the memory to be sufficient for e.g. synchronisation of probabilistic photon sources to generate large optical quantum states at high rates \cite{Nunn2013}. These prospects in conjunction with ORCA's intrinsic compatibility with integrated network architectures \cite{Sprague2014,Kaczmarek2015} render this new protocol a promising candidate for the future up-scaling of photonic quantum networks, opening the way to a new regime of quantum simulation, computation, and sensing.

\begin{acknowledgments}
We would like to thank R. Chrapkiewicz, M. Parniak, M. Beck, S. Gao and J. Sperling for useful discussions. We are grateful to H. Chrzanowski and P. Humphreys for proof-reading the manuscript.

This work was supported by the UK Engineering and Physical Sciences Research Council through Standard Grant No. EP/J000051/1, Programme Grant No. EP/K034480/1, and the EPSRC NQIT Quantum Technology Hub. We acknowledge support from the Air Force Office of Scientific Research: European Office of Aerospace Research and Development (AFOSR EOARD Grant No. FA8655-09-1-3020). J.N. acknowledges a Royal Society University Research Fellowship, and DJS acknowledges an EU Marie-Curie Fellowship No. PIIF-GA-2013-629229. P.M.L. acknowledges a European Union Horizon 2020 Research and Innovation Framework Programme Marie Curie individual fellowship, Grant Agreement No. 705278, and B.B. acknowledges funding from the European Union’s Horizon 2020 Research and Innovation programme under grant agreement No. 665148. I.A.W. acknowledges an ERC Advanced Grant (MOQUACINO). S.E.T. and J.H.D.M are supported by EPSRC via the Controlled Quantum Dynamics CDT under Grants EP/G037043/1 and EP/L016524/1. G.S.T. acknowledges support from the Natural Sciences and Engineering Research Council of Canada (NSERC). E.P. acknowledges an EU Marie-Curie Fellowship No. PIEF-GA-2013-627372. K.T.K. acknowledges a Santander Graduate Scholarship from Lady Margaret Hall, Oxford. O.L.-A. acknowledges Consejo Nacional de Ciencia y Tecnologia (CONACyT) for support from 'Becas Conacyt Al Extranjero 2016' and Banco de M\'exico (BM) for support from 'Fondo para el Desarrollo de Recursos Humanos' (FIDERH).
\end{acknowledgments}

\bibliographystyle{unsrt}

\onecolumngrid
\section*{Appendix}
\setcounter{figure}{0}
\renewcommand{\thefigure}{A.\arabic{figure}}
\setcounter{table}{0}
\renewcommand{\thetable}{A.\arabic{table}}
\setcounter{subsection}{0}
\renewcommand{\thesubsection}{A.\arabic{subsection}}
\setcounter{equation}{0}
\renewcommand{\theequation}{A.\arabic{equation}}

\subsection{Experimental setup \label{setup}}

\begin{figure}[h!]
\includegraphics[width=\textwidth]{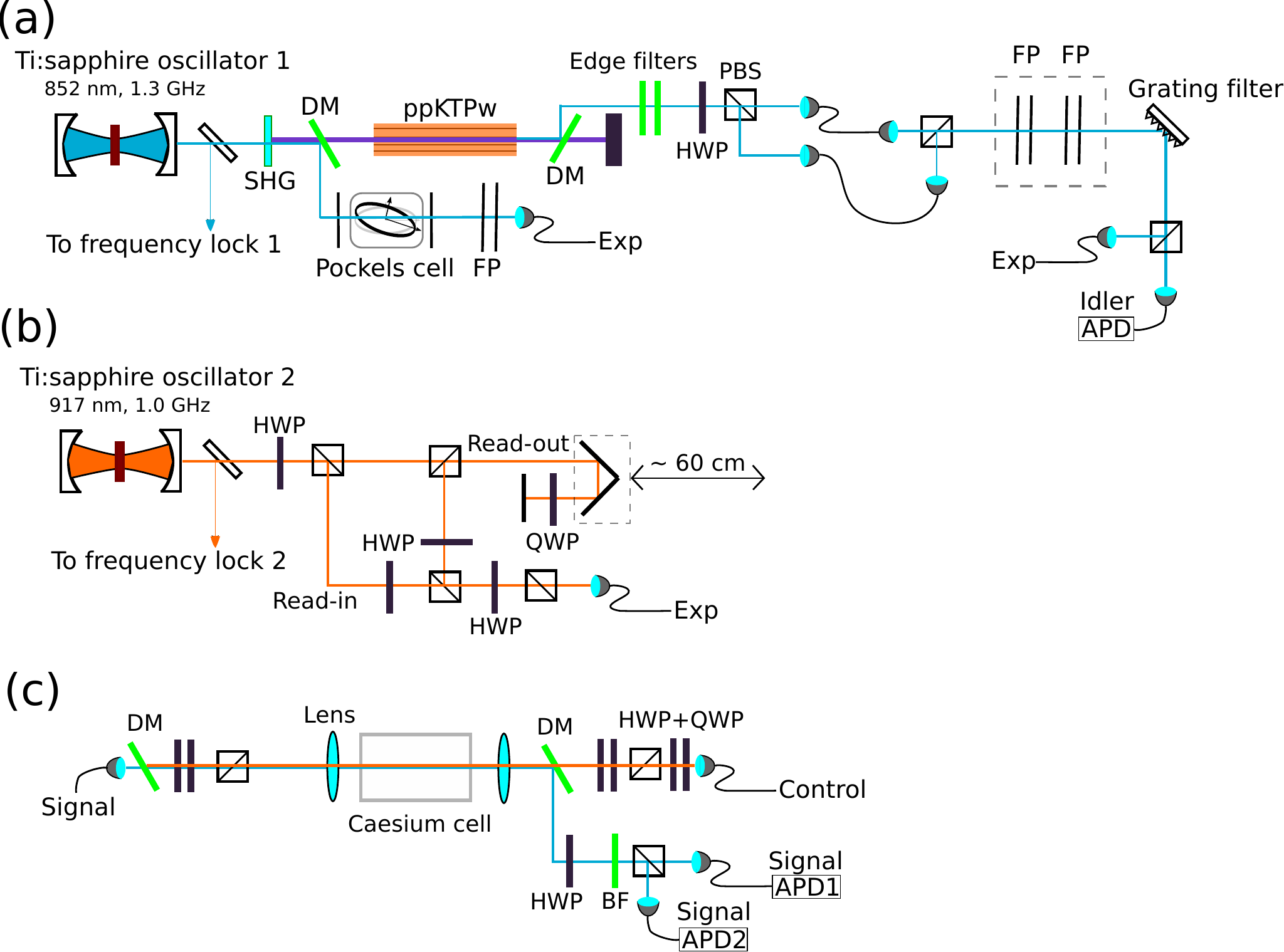} 
\caption{\textbf{Schematic of the experimental setup for ORCA}. \textbf{(a) }  Signal generation stage. \textbf{(b) } Control generation stage. \textbf{(c) } Memory and detection stage. SHG - periodically poled potassium titanyl (ppKTP) bulk crystal; ppKTPw - ppKTP waveguide; DM - dichroic mirror; FP - Fabry-P\'erot etalon; PBS - polarizing beamsplitter; HWP - half-wave plate; QWP - quarter-wave plate; APD - avalanche photodiode detector; BF - bandpass filter.}\label{fig:figSup1}
\end{figure}

Figure \ref{fig:figSup1} shows a schematic of the ORCA experimental setup. For the ``classical'' memory characterisation, we produce a weak coherent state signal (average photon number of $\sim2$) by picking pulses using a fast Pockels cell (extinction ~20,000:1) at a 1 MHz rate from a 80 MHz train of pulses generated by a $\sim330$ ps actively mode-locked titanium sapphire (Ti:sapphire) laser operated at 852 nm and filtered by a Fabry-P\'erot (FP) etalon down to 0.81 GHz bandwidth. Using a scanning FP etalon connected to a PC running LabVIEW, we reference-lock the signal Ti:sapphire's center frequency (via the voltage on a Gires-Tournois-Interferometer inside the laser cavity) to a continuous wave (CW) laser locked to the Cs D2 line via saturated absorption spectroscopy. 

We generate the control field from a second $\sim500$ ps actively mode-locked Ti:sapphire laser operated at 917 nm, with its center frequency locked using a wavelength meter, and its repetition rate locked to the signal Ti:sapphire using a commercial lock-to-clock (L2C) system. We use an unbalanced free-space Mach-Zehnder interferometer to split the 80 MHz pulse train into two, with a variable delay $<\sim4$ ns between them, in order to investigate storage times $<12.5$ ns. For storage times $6\;\mathrm{ns}<\tau<12.5$ ns we use the L2C electronics to change the timing between signal and control pulse trains such that read-in and read-out are switched. We also use the L2C to temporally overlap the signal and control pulses in the memory cell.

We combine the signal and control fields on a dichroic mirror, which - followed by a 10 nm bandpass filter centred at 850 nm - reduces control field leakage to the detectors from back-reflections by a factor of $\sim10^9$. We focus signal and control beams down to a $\sim300\;\mu$m waist inside a 72$\,$mm long uncoated caesium borosilicate reference cell heated using a custom-made oven. We estimate the cell temperature to be $\sim 91^o$C by frequency scanning a weak CW probe laser over the Cs D2 line and fitting a Voigt profile to the measured atomic absorption line.

After the signal field passes through the memory and the filters, we send it into a Hanbury-Brown-Twiss setup, composed of a half-waveplate, polarising beamsplitter and two fibre-coupled single-photon avalanche photodiodes. The two signal and the idler avalanche photodiodes were connected to a time-to-digital converter. For the weak coherent state data and cross-correlation measurements, we add the counts on the two signal detectors to estimate the total magnitude of the transmitted/recalled signal.

\subsection{Memory lifetime \label{lifetime}}
We identify three effects limiting the lifetime of the ORCA memory in our current implementation: spontaneous emission from the doubly excited atomic state, motion-induced dephasing due to the Doppler effect, and oscillations due to quantum interference in the doubly excited state manifold.

Motion-induced dephasing arises due to the Doppler effect from atomic motion in a warm ensemble. This is because the stored excitation is spread over atoms belonging to different velocity classes in the ensemble. Each velocity class experiences Doppler shifted frequencies for the signal and control fields. As a consequence, the phase of the stored coherence evolves at different rates in different velocity classes. The motion-induced dephasing lifetime is $\tau_D = \frac{1}{k_r v_s}$ [27], where $v_s = \sqrt{k_B T/m}$, and $k_r=\frac{2\pi}{\lambda_\mathrm{s}} -\frac{2\pi}{\lambda_\mathrm{c}}$; with $T$ the temperature of the atomic vapour, $m$ the mass of the atom, and $\lambda_\mathrm{s/c}$ the wavelengths of the signal and control fields. In other words, the collective coherence will dephase at a rate proportional to the square root of the temperature, and the wavenumber mismatch of the fields.

\begin{figure}[h]
\centering
\includegraphics[width=\linewidth]{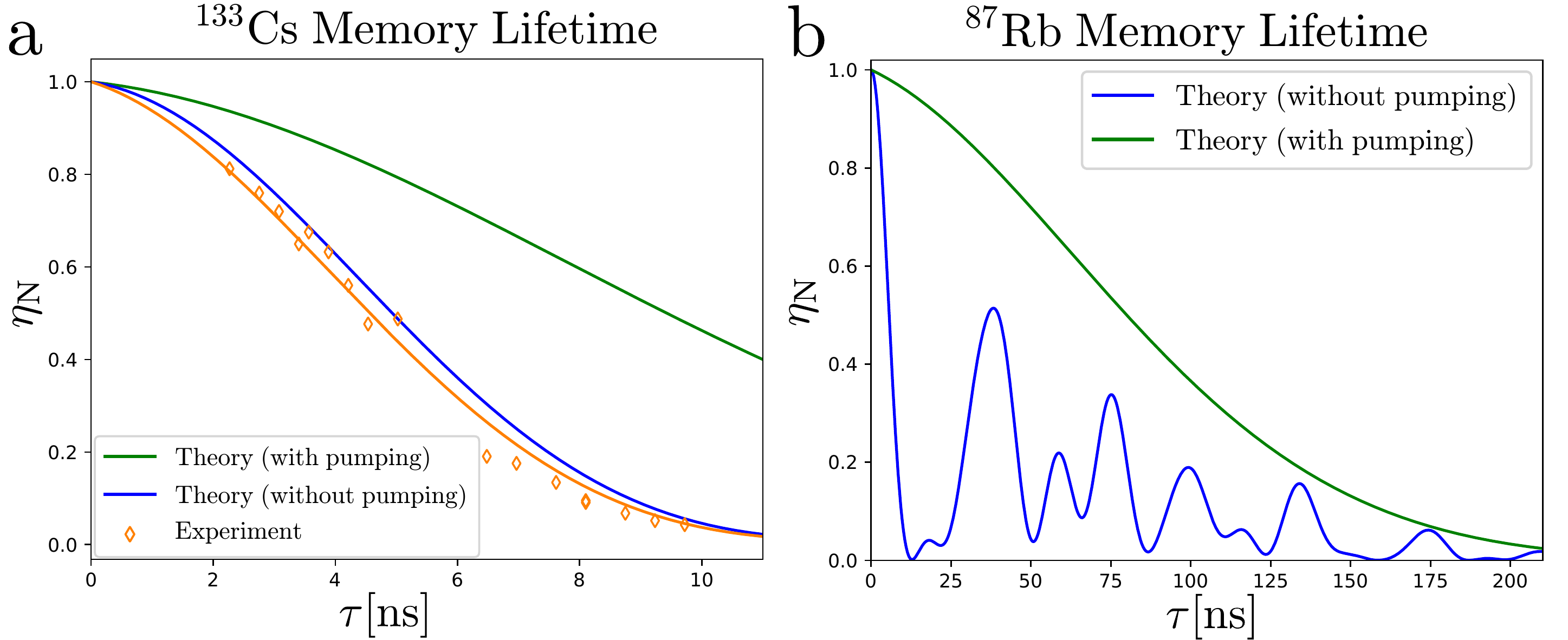}
\caption{\textbf{a}. The measured (normalised to $\tau=0$) memory efficiency $\eta_\mathrm{N}$ (orange diamonds, experimental errors are smaller than the markers) versus storage time $\tau$ along with a theory fit of the memory lifetime curve (orange line) yielding a (1/e) lifetime of $5.4\pm0.1$ ns. Also shown is the predicted memory lifetime curves with (green) and without (blue line) quantum interference in the doubly excited storage state; possible via optical pumping prior to memory operation. \textbf{b}. Memory lifetimes predicted from theory for $^{87}\mathrm{Rb}$ with (green) and without (blue line) similar quantum interference.}
\label{fig:figLifetime}
\end{figure}

In the absence of optical pumping, the broadband two-photon excitation that stores the signal has contributions from all allowed paths connecting the $6S_{1/2}$ and $6D_{5/2}$ manifolds. The resulting excitation is thus spread across the different hyperfine components of the $6D_{5/2}$ manifold. During storage these components oscillate with different rates as given by their energy separations, and at read-out they can interfere destructively (especially visible in Fig. \ref{fig:figLifetime}b). Optical pumping restricting the memory interaction to the hyperfine levels $6S_{1/2}(F=4) \rightarrow 6P_{3/2}(F=5) \rightarrow 6D_{5/2}(F=6)$ would reduce this effect [37].

We determine the actual memory lifetime by measuring the memory efficiency for different storage times using a weak coherent state signal. In Fig. \ref{fig:figLifetime}a we show the measured normalized memory efficiency versus storage time (orange diamonds). Fitting our model to the data we obtain a $1/e$ lifetime of $5.4\pm 0.1$ ns. Using the Maxwell-Bloch model (which includes motion-induced dephasing, as well as the hyperfine oscillations) we predict a memory lifetime of 5.9 ns, very close to the measured one (blue line in Fig. \ref{fig:figLifetime}a).

The memory lifetime can be extended through optical pumping to reduce the destructive interference of hyperfine components and/or by using a different atomic species with a smaller signal/control wavenumber mismatch. We can model the effect of optical pumping by neglecting dipole couplings in the Maxwell-Bloch model such that only the transition $6S_{1/2}(F=4) \rightarrow 6P_{3/2}(F=5) \rightarrow 6D_{5/2}(F=6)$ is allowed. In this way we obtain a memory lifetime of 11.5 ns (the green curve in Fig. \ref{fig:figLifetime}a). Furthermore, a simulation of the $5S_{1/2}\rightarrow5P_{3/2}\rightarrow5D_{5/2}$ cascade in $^{87}\mathrm{Rb}$ (signal at 780~nm, control at 776~nm) yields a memory lifetime of $99$~ns as shown by the green curve in Fig. \ref{fig:figLifetime}b. Indeed, recently Finkelstein et al. demonstrated a fast ladder memory (FLAME) --- equivalent to ORCA when far-detuned --- using classical pulses in this system and showed a lifetime of around 85 ns \cite{Finkelstein2017}. 

\subsection{Photon source and setup losses}
The generation of heralded single photons is achieved using type-II parametric down-conversion in a periodically poled potassium titanyl (ppKTP) waveguide. The waveguide, operated in a single-pass configuration and of length 20$\,$mm, is pumped with pulses of approximately 270 ps duration at a wavelength of 426\,nm. This pump light is derived by doubling the above mentioned 852\,nm Ti:sapphire laser via second harmonic generation in a separate 2$\,$mm long ppKTP crystal. With an incident average power near 700\,mW at 852\,nm at the crystal we arrive with 4\,mW average power at 426\,nm before the PDC waveguide. This light is then coupled to the waveguide with a total transmission of $<10\%$ including the loss at the in- and out-coupling lenses. We note that the waveguide is not single-mode for the pump wavelength and that the coupling is optimised to primarily excite the fundamental spatial mode, resulting in a low overall transmission. The generated frequency-degenerate but polarisation-orthogonal signal and idler modes have a bandwidth on the order of $1\,$THz. 

We characterise beam propagation transmission using an ``alignment'' mode, which is coupled to the fundamental mode of the waveguide, and thereby comparable to the signal and idler modes allowing for ``classical'' measurements to be made. These modes are then subject to frequency filtering. First, we apply coarse filtering using edge filters. Then, the modes are spatially separated via a PBS to then be coupled to their own single-mode fibre (SMF) with an efficiency of $(64 \pm 1)\%$ for the signal mode and $(53 \pm 2)\%$ for the idler mode. The modes are then out-coupled and recombined on a PBS forming a common spatial mode to then pass two etalons, one of FSR$\,$=$\,$18.4$\,$GHz and one of $103\,$GHz, which gives an effective FSR of $~1\,$THz (lowest common multiple). This is followed by a holographic volume Bragg grating with a width $\sim$100$\,$GHz. Using a narrowband ($\sim\,$MHz) probe the measured width this filtering has is $1\,$GHz for both modes. Finally the modes are separated spatially again via a PBS, the idler coupled to an APD (efficiency $\eta \approx 50\%$, dark counts = $163\pm 1\,$Hz) via a multimode fibre (total transmission from after waveguide to in front of idler detector is $\eta_\textrm{i,filt}\,=\,(9.7\pm0.1)\,\%$), while the signal is coupled to a SMF to be out-coupled and steered to the memory (total transmission from after waveguide to after this SMF is $\eta_\textrm{s,filt}\,=\,(12.8 \pm 0.3)\,\%$). 

The filtered signal photon is now steered toward the memory. First there is an edge filter, which is used to prevent the 917$\,$nm control from backward-coupling toward the source, which presents additional loss to the signal mode. Further, the caesium cell used is uncoated, adding more loss. After passing the cell, the signal mode is then separated from the control mode via a dichroic mirror and finally passes a bandpass filter centered about 852$\,$nm before entering a Hanbury-Brown-Twiss set-up. The mode is spatially separated into two and coupled to two APDs ($\eta\approx50\,\%$ dark counts = $296 \pm 2\,$Hz and $\eta\approx50\,\%$ dark counts = $356 \pm 2\,$Hz) via SMF. The total transmission from the source to in front of these detectors (averaging over the two SMF couplings) is $\eta_\textrm{s,total}\,=\,3.7\pm0.1\,\%$. That is to say, the photon undergoes an additional $\eta_\textrm{s,add}\,=\,30\,\%$ transmission from after the initial filtering stage.

For all results presented in this paper we operated with an average pump power of $4\,$mW in front of the waveguide in-coupling lens. Typically, we measure an idler (signal) count rate of around $30\,$kHz ($10\,$kHz) for the case of no control pulses; limited by current filter losses. The typical Klyshko efficiency $\eta_k$ measured is $0.7\,\%$. This allows to calculate a heralding efficiency in front of the memory to be $\eta_\textrm{herald} = \eta_k/\eta_{\textrm{det}}/\eta_\textrm{s,add} = 4.7\,\%$, which is well above the $\mu_1$ of the memory, as required for single-photon storage [12]. Finally, the heralding efficiency just after the waveguide is $\eta_\textrm{s,waveguide} = \eta_k/\eta_{\textrm{det}}/\eta_\textrm{s,total} = 38\,\%$. The missing factor of $2.6$ we attribute to not measuring explicitly the loss inside the waveguide, the out-coupling loss from waveguide to free-space and the potential frequency mismatch of the etalon pass bands between signal and idler.

\subsection{Data acquisition and post-processing}
\begin{figure}[h!]
\centering
\includegraphics[width=\linewidth]{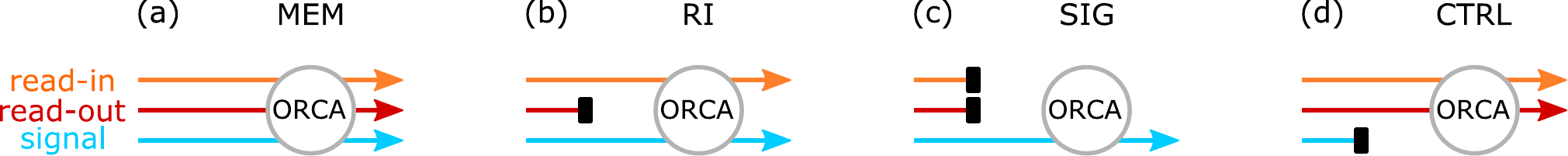}
\caption{\textbf{Schematic of the different measurement settings.} Black rectangles signify a closed shutter. For more information see the text.}
\label{fig:measurement_settings}
\end{figure}

During the measurements, the settings of three mechanical shutters which selectively blocked the read-in, read-out, and signal beams defined four different configurations (see Fig. \ref{fig:measurement_settings}): memory measurements with all three shutters open (``MEM''); read-in measurements with signal and read-in shutters open, and read-out shutter closed (``RI''); signal measurements with signal shutter open and both read-in and read-out closed (``SIG''); and noise measurements with read-in and read-out shutters open and signal shutter closed (``CTRL''). A single measurement consisted of recording the number of detector counts registered in a period of 180~s in the ``MEM'' configuration, followed by recording the total counts over 10~s in the ``RI'', ``SIG'', and ``CTRL'' configurations. After completion of all four configurations, the corresponding data was written to disk and the measurement repeated. This mode of operation was chosen in order to mitigate the effect of slow drifts in the setup that arose from changes in the laboratory temperature and humidity.

For each configuration in each measurement, we recorded arrival time histograms for the three detectors ($\mathrm{D_{i}}$, $\mathrm{D_{s1}}$, $\mathrm{D_{s2}}$). These are histograms of firing times of the single-photon detectors with respect to a 1~MHz trigger signal derived from the Ti:sapphire laser recorded with the time-to-digital converter (TDC). We chose a time-bin width of 200~ps as a compromise between temporal resolution of the TDC and total number of time bins in the histogram. For data visualization, we added all arrival time histograms and normalised them to both the number of measurements (521) and the respective measurement duration (180~s for ``MEM'', 10~s else), obtaining a count rate per time bin in units of Hertz. 

\begin{figure}
\centering
\includegraphics[width=\linewidth]{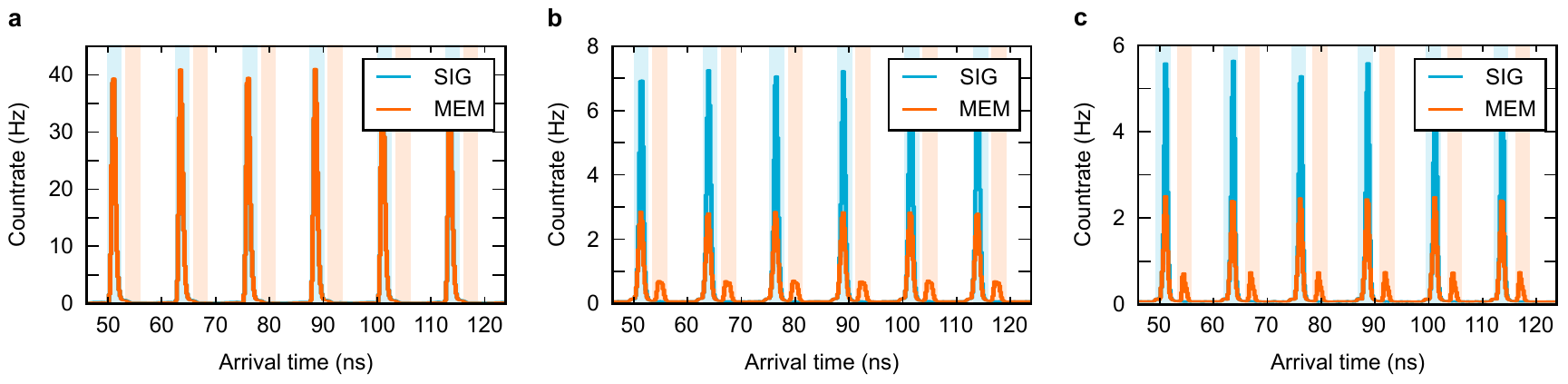}
\caption{\textbf{Sections of the arrival time histograms with indicated time gates.} \textbf{a} A section of the arrival time histogram for $\mathrm{D_{i}}$ We show the histograms for the ``SIG'' (blue trace) and ``MEM'' (orange trace) configuration. \textbf{b} A section of the arrival time histograms for detector $\mathrm{D_{s1}}$. \textbf{c} The same for detector $\mathrm{D_{s2}}$. For more details see the text.}
\label{fig:arrival_time_02}
\end{figure}

Fig. \ref{fig:arrival_time_02} shows a section of the arrival time histograms. To reduce the impact of spurious noise counts (primarily from detector dark counts), we applied time gates to the recorded arrival time histograms and only kept events that lay within the time gates. The time gates for the read-in pulses (blue regions) are centred around the maxima of the individual read-in peaks and have a width of 2.5 ns, chosen such that the peaks were completely inside the gating region. Similar time gates were chosen for the recalled light, where the centre of these read-out gates (orange regions) was offset from the corresponding read-in time gates by 3.5 ns, which was the storage time chosen for the experiment. By integrating the detection events over only the gate regions, we calculated the read-in, read-out, and total memory efficiencies stated in the main text.  

In addition to the arrival time histograms, we also recorded coincidence histograms for signal-idler two-fold coincidences ($\mathrm{D_i}$\&$\mathrm{D_{s1}}$, $\mathrm{D_i}$\&$\mathrm{D_{s2}}$) and the three-fold coincidences ($\mathrm{D_i}$\&$\mathrm{D_{s1}}$\&$\mathrm{D_{s2}}$). These are start-stop histograms, where the detection of an idler photon starts the measurement and the detection of a signal photon (the detection of a $\mathrm{D_{s1}}$\&$\mathrm{D_{s2}}$ coincidence) serves as the stop signal for the two-fold (three-fold) coincidence measurement. Here, the time-bin width of the TDC was chosen to be 100~ps to ensure that the temporal resolution of the measurement was not limited by the TDC, and the time gates had a width of 3.5 ns. Again, the data was post-processed for visualization similar to the arrival time histograms. The resulting coincidence traces are plotted in Fig. S.5. Note that the unconventional shape of the traces originates from the logarithmic scaling of the y axes.

\begin{figure}
\centering
\includegraphics[width=\linewidth]{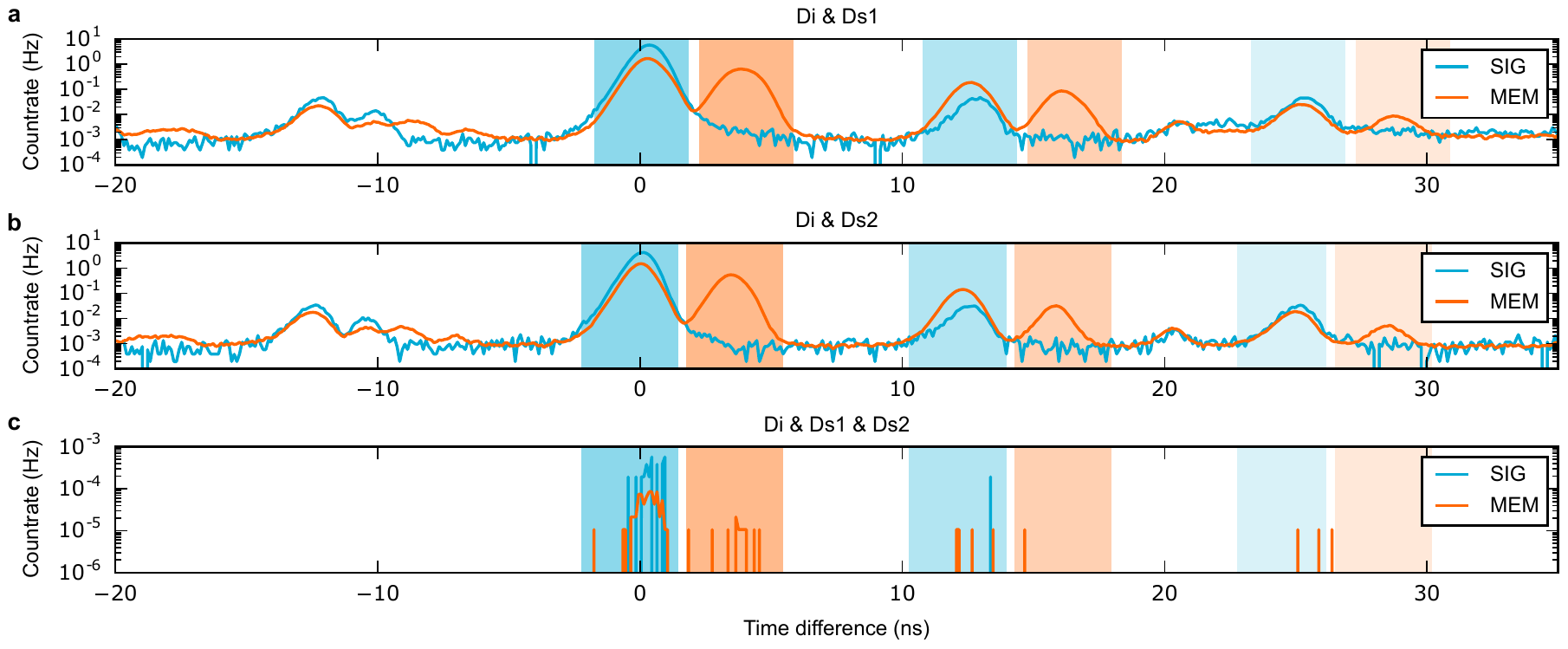}
\caption{\textbf{Correlation histograms.} \textbf{a} Accumulated normalised correlation histogram for two-fold coincidences between detectors $\mathrm{D_{i}}$ \& $\mathrm{D_{s1}}$, shown for the ``SIG'' (blue trace) and ``MEM'' (orange trace) configurations. Note the logarithmic scaling of the y axis. In the main text, we analyse the $g^{(1,1)}(0)$ cross-correlation for the initial time at 0~ns (read-in; blue-shaded) and 3.5~ns (read-out; orange-shaded). Successive read-in (read-out) time bins are indicated by shaded regions with decreasing saturation. \textbf{b} The same as in \textbf{a}, now however for two-fold coincidences between detectors $\mathrm{D_{i}}$ \& $\mathrm{D_{s2}}$. \textbf{c} Correlation histogram for three-fold coincidences between detectors $\mathrm{D_{i}}$ \& $\mathrm{D_{s1}}$ \& $\mathrm{D_{s2}}$.}
\label{fig:correlations}
\end{figure}

The ``SIG'' traces show a dominant peak at a time difference of 0~ns, with subsequent smaller peaks at integer multiples of the laser repetition time of 12.5~ns. From this we calculate the $g^{(1,1)}$ signal-idler cross-correlation function. In order to do so, we use

\begin{equation}
g^{(1,1)}=\frac{R_\mathrm{s,i}}{R_\mathrm{s}R_\mathrm{i}}R_\mathrm{T},
\end{equation}
where $R_\mathrm{s,i}$ is the sum of $\mathrm{D_{i}}$\&$\mathrm{D_{s1}}$ and $\mathrm{D_{i}}$\&$\mathrm{D_{s2}}$ coincidences, $R_\mathrm{T}$  is the total number of trigger events during the whole measurement time, $R_\mathrm{s}$ is the sum of $\mathrm{D_{s1}}$ and $\mathrm{D_{s2}}$ clicks, and $R_\mathrm{i}$ is the number of $\mathrm{D_{i}}$ clicks.

The results for the ``SIG'' configuration are summarized in the first row of Tab. S.1, where we find $g^{(1,1)}=130(5)$ for a time difference of 0~ns and $g^{(1,1)}\approx1$ for integer multiples of 12.5~ns. We also note that the values at the read-out times (3.5~ns offset from the 12.5~ns time slots) are meaningless, since there is no actual signal at the detectors. 

Turning our attention to the ``MEM'' configuration (orange traces in Fig. S.5), we again find a dominating peak at a time difference of 0~ns with side peaks at integer multiples of 12.5~ns. In addition, we see peaks that are offset from the major peaks by 3.5~ns. These originate from coincidence events between idler photons and signal photons that have been stored in and recalled from the memory. We also note that the side peak at 12.5~ns is higher than the corresponding peak for the``SIG'' configuration. The reason for this lies in the non-unity read-out efficiency of our memory. A stored photon is not necessarily read out after 3.5~ns, but can remain stored in the memory. Then, it can be read out by the next laser pulse arriving at 12.5~ns, and so on. To quantify this effect, we again evaluate $g^{(1,1)}$. The results are summarized in the second row of Tab. S.1. In this case, we find non-classical values for $g^{(1,1)}$ up to a time of 16~ns, which corresponds to around three times the lifetime of our memory. These results highlight the noise-free operation of ORCA: non-classical photon correlations are retained even after the memory efficiency has decayed to around 5\% of its initial maximum value ($1/e^3$).

\begin{table}
\centering
\begin{tabular}{|l||l|l|l|l|l|l|}
\hline
\multirow{2}{*}{}   & \multicolumn{6}{c|}{$g^{(1,1)}$} \\
    & $\tau=0$ ns & $\tau=3.5$ ns & $\tau=12.5$ ns & $\tau=16$ ns & $\tau=25$ ns & $\tau=28.5$ ns \\
\hline\hline
SIG & 130(5) & ------ & 1.1(2) & ------ & 1.2(2) & ------ \\
\hline
MEM & $80(3)$ & $120(5)$ & $9.7(4)$ & $11.3(5)$ & $1.7(2)$ & $1.1(1)$ \\
\hline
\end{tabular}
\caption{\textbf{Cross-correlation for successive read-outs.} Calculating $g^{(1,1)}$ for higher-order read-outs at integer multiples of $12.5$ ns (plus 3.5 ns for read-out pulses) yields the preservation of non-classical correlations by the memory up to around three times the memory lifetime.}\label{tab:g11}
\end{table}

As for the heralded auto-correlation $g^{(2)}_\mathrm{h}$, we evaluate it from the measurements using

\begin{equation}
g^{(2)}_\mathrm{h}=\frac{R_\mathrm{trip}}{R_\mathrm{s1,i}R_\mathrm{s2,i}}R_\mathrm{i},
\end{equation}
where $R_\mathrm{trip}$ is the number of triple coincidences between $\mathrm{D_{i}}$\&$\mathrm{D_{s1}}$\&$\mathrm{D_{s2}}$, $R_\mathrm{i}$ is the number of idler clicks, and $R_\mathrm{s1(2),i}$ is the number of $\mathrm{D_{i}}$\&$\mathrm{D_{s1}}$ ($\mathrm{D_{i}}$\&$\mathrm{D_{s2}}$) coincidences.

\end{document}